\documentclass[prb,aps,twocolumn,preprintnumbers,10pt,floats,amsmath,showpacs,amssymb,superscriptaddress]{revtex4-1}

\usepackage{hyperref}
\usepackage{dcolumn}
\usepackage{bm}
\usepackage{amssymb}
\usepackage{subfigure}
\usepackage{color}
\usepackage{amsmath}
\usepackage{graphicx}
\usepackage{hyperref}
\usepackage{psfrag}
\usepackage{color} 
\usepackage{simplewick}
\usepackage{appendix}
\usepackage{color}
\usepackage{comment}
\usepackage{bbold}
\usepackage{mathrsfs}

\DeclareMathOperator{\e}{e}

\allowdisplaybreaks

\def \be{\begin{equation}}
\def \ee{\end{equation}} 

{\left\lbrace\begin{array}{@{}l@{}}}%
{\end{array}\right.}

\begin{document}

\title{Separation of heat and charge currents  for boosted thermoelectric conversion}

\author{Francesco Mazza}
\affiliation{NEST, Scuola Normale Superiore, and Istituto Nanoscienze-CNR, I-56126 Pisa, Italy}

\author{Stefano Valentini}
\affiliation{NEST, Scuola Normale Superiore, and Istituto Nanoscienze-CNR, I-56126 Pisa, Italy}

\author{Riccardo Bosisio}
\affiliation{SPIN-CNR, Via Dodecaneso 33, 16146 Genova, Italy}
\affiliation{NEST, Istituto Nanoscienze-CNR and Scuola Normale Superiore, I-56126 Pisa, Italy}

\author{Giuliano Benenti}
\affiliation{Center for Nonlinear and Complex Systems, Universit\`a degli Studi dell'Insubria, 
		via Valleggio 11, 22100 Como, Italy}
\affiliation{Istituto Nazionale di Fisica Nucleare, Sezione di Milano, via Celoria 16, 20133 Milano, Italy}

\author{Vittorio Giovannetti}
\affiliation{NEST, Scuola Normale Superiore, and Istituto Nanoscienze-CNR, I-56126 Pisa, Italy}

\author{Rosario Fazio}
\affiliation{NEST, Scuola Normale Superiore, and Istituto Nanoscienze-CNR, I-56126 Pisa, Italy}

\author{Fabio Taddei}
\affiliation{NEST, Istituto Nanoscienze-CNR and Scuola Normale Superiore, I-56126 Pisa, Italy}


\begin{abstract}
In a multi-terminal device the (electronic) heat and charge currents can follow different paths. In this paper we introduce and analyse a class 
of multi-terminal devices where this property is pushed to its extreme limits, with charge {\em and} heat currents flowing in different reservoirs.
After introducing the main characteristics of such {\em heat-charge current separation} regime we show how to realise it in a multi-terminal 
device with normal and superconducting leads.
We demonstrate that this regime allows to control independently heat and charge flows and to greatly enhance thermoelectric performances at low temperatures.  
We analyse in details a three-terminal setup involving a superconducting lead, a normal lead and a voltage probe. For a generic scattering region we show 
that in the regime of heat-charge current separation both the power factor and the figure of merit $ZT$ are highly increased with respect to a 
standard two-terminal system.
These results are confirmed for the specific case of a system consisting of three coupled quantum dots.
\end{abstract}


\maketitle

\section{Introduction}
Increasing the efficiency of thermoelectric materials for heat-work conversion is one of the main challenges of present-days 
technology~\cite{Mahan1997,Majumdar2004,Dresselhaus2007, Snyder2008,Shakouri2011,Dubi2011}. In this context the 
search for efficient nanoscale heat engines and refrigerators has stimulated a large body of activity, recently reviewed in Ref.~\onlinecite{Benenti2013}. 
Progresses in understanding thermoelectricity at the nanoscale will also have important applications for 
ultra-sensitive all-electric heat and energy transport detectors, energy transduction, heat rectifiers and refrigerators, just to 
mention a few examples.
One of the keys to the success in this field is the ability to modulate, control, and route heat and charge currents, ideally 
achieving their separate control~\cite{Jian2013,Imry2015,Jian2014a,Jian2014b, Jordan2013, Sothmann2013, Sothmann2014}.
This is however by no means obvious as the charge and (the electronic contribution to)
the heat are transported by the same carriers. In two-terminal systems, for example, within the linear response regime, electrical and thermal currents are strictly 
interrelated, as manifested by the emergence at low enough temperatures of the Wiedemann-Franz law~\cite{ashcroft1976}. 
Indeed, when the temperature is the smallest energy scale in the system (that is, if the Sommerfeld expansion holds) one finds 
that the ratio $\Lambda = K/(G T)$, involving the $electrical$ $G$ and the $thermal$ $K$ conductances at temperature $T$ is universal and it is given by 
the Lorenz number $\Lambda_0=\pi^2/3\,\left(k_B/ e\right)^2$. The fulfillment of the Wiedemann-Franz law, one of the triumphs 
of Sommerfeld's theory of metals, has important consequences in determining the efficiency of thermoelectric engines. For a two 
terminal setup the only way to increase the thermoelectric figure of merit $ZT=(G{\cal S}^2/K)T$, the dimensionless parameter that fully describes 
the efficiency for thermoelectric conversion, is by increasing the thermopower (${\cal S}$), which is however bounded to be small 
at small $T$~\cite{Benenti2013}. 
\begin{figure}[!h]
\centering
\includegraphics[width=\columnwidth,  keepaspectratio]{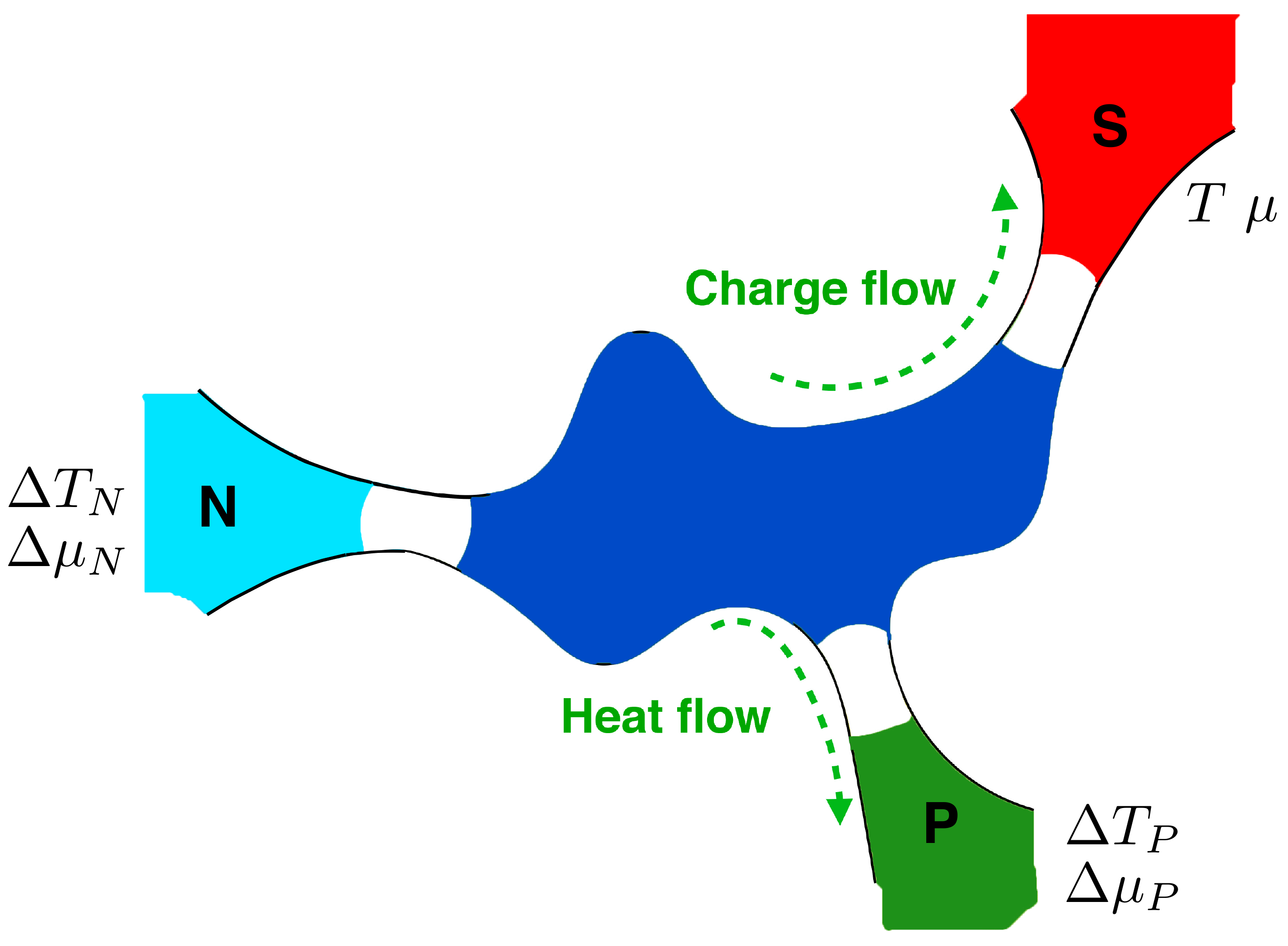}
\caption{The heat-charge current separation scheme. A generic scattering region is connected to three reservoirs labeled by the letters S (superconducting lead), P (voltage probe) and N (normal metal lead). 
In the main text we assume the superconducting reservoir (S) to be the reference. In any case, as pointed by the arrows, only charge flows inside lead S whereas only heat flows inside lead P.} 
\label{fig:SPN}
\end{figure}

In order to achieve a separate control of heat and charge currents one should therefore consider more complex (multi-terminal) devices. 
The key issue in this context is to assess to which extent this control can be achieved and what are its possible advantages in thermoelectric 
thermal machines. In this paper we push to its extreme this type of control and explore a situation where heat and charge currents
flow in spatially separated parts of the system.
Namely, we will enforce that one of the terminals allows only charge current and another one allows only heat current, and name this
regime as {\em Heat-Charge Current Separation} (HCCS).
It is worth stressing already at this point 
that the regime of HCCS is realised notwithstanding the fact that the same carriers are responsible for heat and charge flow. As we will 
discuss in details in the paper, HCCS can be naturally realised by employing superconducting reservoirs. Nonetheless, this is not a strict requirement: 
indeed, in principle one could spatially separate heat and charge currents in an all-normal multi-terminal device. 
In this case, however, a fine tuning of the parameters characterising the thermoelectric transport needs to be performed thus making 
the device not easy to realise experimentally.

Recent investigations of multi-terminal setups~\cite{Balachandran2013,Entin-Wohlman2010,Entin-Wohlman2012,
Horvat2012,Jacquet2009, Bedkihal2013,Bosisio2014,Bosisio2014(b),Brandner2013, 
Brandner2013(b),Jiang2012,Jiang2013,Saito2011,Sanchez2011,Sanchez2011(b),Sothmann2012,Sothmann2012(b), Amikam2014} have shown that 
such devices offer great potentialities in terms of efficient thermoelectric conversion. In the majority of these works all but two terminals were 
considered as probes, i.e. no net flow of energy and charge through them was allowed. In other works a purely bosonic reservoir was added to the standard two terminals, only exchanging energy 
with the system.  A generic three-terminal setup, where all reservoirs are fermionic (possibly exchanging both charges and heat with the 
system), was considered in Ref.~\onlinecite{Mazza2014} where it was shown that the third terminal can be used to increase both the extracted 
power and the efficiency of a thermal machine. So far the impact of superconducting reservoirs on the performance of thermoelectric devices has 
not been considered~\cite{thpowersup}. We start filling this gap by studying HCCS and thermoelectric conversion in  a three-terminal 
hybrid normal metal-superconducting device.

The three-terminal device which implements HCCS, pictorially shown in Fig.~\ref{fig:SPN}, is composed of a generic conductor connected to a superconducting reservoir (S), a normal metal reservoir (N) and a second normal reservoir whose chemical potential is set to inhibit the flow of electrical current, thus acting as a voltage probe (P).
This setup, to which we will refer to as SPN, \emph{naturally} realises heat-charge current separation.
Indeed, a voltage probe exchanges (on average) by definition only heat (energy) with the system, whereas the superconductor, being a poor heat 
conductor for temperatures below the gap, can exchange only charges. This way, the heat and charge currents, flowing together out 
of the normal metal reservoir (N), are split and driven either towards the voltage probe (heat), or towards the superconducting reservoir (charge). 
In the linear response regime this setup has the advantage of admitting an effective description in terms of a $2\times 2$ Onsager matrix, a 
feature which allows \emph{inter alia} a natural way of comparing its performance to that of a standard two-terminal configuration. Using 
the scattering approach, we will show on general grounds that this separation allows, in the linear response regime and at small temperatures,
to greatly enhance the performance of a thermal machine, namely increasing both the efficiency and the output power by roughly one order 
of magnitude with respect to a standard two-terminal counterpart. The root of this enhanced efficiency can be traced back to the possibility 
to violate in a controlled fashion the Wiedemann-Franz law in the heat-charge separation regime. On more general grounds it is worth to 
stress that the simultaneous presence of superconducting {\em and} normal terminals, by selectively controlling the heat and charge flows through normal and 
Andreev scattering, introduces new degrees of freedom that are worth being explored for thermoelectric conversion.

The paper is organized as follows: in Section~\ref{sec:model} we introduce the necessary formalism and define the regime of  heat-charge 
current separation. We then show how this regime can be attained by having one of the three terminals in the superconducting state. In order
to test the performance of this thermal engine we perform an extensive analysis in Section~\ref{sec:results}, by varying the properties of the 
scattering region connecting the three reservoirs. By properly parametrising the scattering matrix we sample randomly the scatterer and 
compare the efficiency of the HCCS thermal machine with that of a ``conventional" two-terminal setup (Sections~\ref{sec:violation} and 
~\ref{sec:performance}). We complete our analysis in Section~\ref{sec:dots} by discussing the case of systems consisting of quantum dots (QDs).
The reason to study these examples in detail is to show that it is possible to achieve, in experimentally realisable situations, those enhanced 
performances that we found in the first part of Section~\ref{sec:performance}. Indeed we see that our theoretical findings 
can be tested with current experimental capabilities. Section~\ref{conclusions} is devoted to the concluding remarks. 
Some technical details related to the scattering formalism in the presence of Andreev scattering are summarised in the Appendices.

\section{Heat-charge current separation}
\label{sec:model}

Let us consider a system composed of a conductor attached to three leads. Within the linear response regime charge and 
heat currents are governed by the Onsager matrix $\bf{L}$ via the relation 
\begin{equation}\label{eq::onsag_full}
\begin{pmatrix}
J^c_N \\
J^h_N \\
J^c_P\\
J^h_P \\
\end{pmatrix}
= 
\begin{pmatrix}
L_{11} & L_{12} & L_{13} & L_{14} \\
L_{21} & L_{22} & L_{23} & L_{24} \\
L_{31} & L_{32} & L_{33} & L_{34} \\
L_{41} & L_{42} & L_{43} & L_{44} \\
\end{pmatrix}
\begin{pmatrix}
 X^\mu_N \\
 X^T_N \\
 X^\mu_P \\
 X^T_P
\end{pmatrix},
\end{equation}
where $J^c_i$ ($J^h_i$) represent the charge (heat) current entering the conductor from lead $i$, with $i=$ (N, P), see Fig.~\ref{fig:SPN}. 
We define the biases $X^\mu_i = \Delta \mu_i/T = (\mu-\mu_i)/T$ and $X^T_i = \Delta T_i / T^2 = (T- T_i)/T^2$, where $\mu_i$ and 
$T_i$ are the chemical potential and temperature, respectively, relative to reservoir $i=$N,P and having chosen the reservoir S as 
reference with temperature $T$ and chemical potential $\mu$. Heat and charge currents flowing in lead S can be determined from the conservation of particle and energy currents.

As already mentioned,  HCCS consists in spatially separating heat and charge flows. In the example of  Fig.~\ref{fig:SPN} heat will only flow 
in lead P while charge will only flow in lead S. In this section we characterise this regime and discuss how to implement it.

On general grounds HCCS can be realised whenever two ``probe" terminals~\cite{Gramespacher1997} are present, one for the voltage and one for the 
temperature. In fact, a voltage probe is a terminal whose voltage is adjusted in order for the charge current to vanish, while a temperature probe is 
a terminal whose temperature is adjusted in order for the heat current to vanish (see App.~\ref{app-VP}).
Unlike a voltage probe, which is implemented simply by opening the electric circuit, making a thermal probe would 
require the ability to control and measure heat currents which is still very challenging in practice (although important advancements in the measurements of heat currents at the nanoscale have been recently achieved, see Refs.~\onlinecite{blanc:2013, meier:2014}).
A natural way of realising HCCS is to replace the thermal probe with a superconducting lead which intrinsically 
suppresses the heat flow for low enough voltages and temperatures. On the contrary a normal metal-superconductor interface is an 
excellent electrical conductor due to the Andreev process that allows to carry charge current in the sub-gap regime.  In the following we 
will detail the working principles of this implementation. 

Let us consider Eq.~\eqref{eq::onsag_full} and take the superconducting reservoir as the reference.
Assuming temperatures much smaller than the superconducting gap and using the scattering formalism (see App.~\ref{app:LBsc}) one can demonstrate 
that the coefficients on the fourth row (column) of the Onsager matrix Eq.~\eqref{eq::onsag_full} are the 
opposite of the corresponding coefficients on the second row (column). This implies that $J_N^h = - J_P^h$ which, at first order in linear response, 
yields $J_S^h=0$ by virtue of the energy conservation. In other words, it is an intrinsic property of the hybrid scattering 
matrix to have vanishing heat current in the superconducting lead. These observations allow us to simplify the Onsager system of equations 
by eliminating the redundant forth row and column, thus reducing it to a 3 by 3 problem:
\begin{equation}\label{eq::onsag_3x3}
\begin{pmatrix}
J^c_N \\
J^h_N \\
J^c_P
\end{pmatrix}
= 
\begin{pmatrix}
L_{11} & L_{12} & L_{13} \\
L_{21} & L_{22} & L_{23} \\
L_{31} & L_{32} & L_{33} \\
\end{pmatrix}
\begin{pmatrix}
X^\mu_N \\
X^T \\
X^\mu_P
\end{pmatrix},
\end{equation}
where we have introduced $ X^T=X^T_N - X^T_P$ (or equivalently $\Delta T = \Delta T_N - \Delta T_P$). Now we impose the voltage probe condition $J^c_P=0$ on reservoir P, which yields
\begin{equation}\label{eq::voltage_probe_supercond}
X^\mu_P = - \frac{L_{31} X^\mu_N + L_{32} X^T}{L_{33}}.
\end{equation}
By substituting Eq.~\eqref{eq::voltage_probe_supercond} into Eq.~\eqref{eq::onsag_3x3} one obtains a two-terminal-like Onsager matrix:
\begin{equation}\label{eq::onsag_sc_probe}
\begin{pmatrix}
J^c_N \\
J^h_N
\end{pmatrix}
= 
\begin{pmatrix}
L'_{11} & L'_{12} \\
L'_{21} & L'_{22} 
\end{pmatrix}
\begin{pmatrix}
X^\mu_N \\
X^T
\end{pmatrix}.
\end{equation}
For the sake of simplicity, in the following we drop the primes for the Onsager coefficients $L'_{ij}$ in Eq.~\eqref{eq::onsag_sc_probe}.
From the definitions of the local~\cite{Benenti2013} and non-local~\cite{Mazza2014} transport coefficients, one can introduce 
(local) conductances and (non-local) thermopowers described by the following two-terminal-like expressions:
\begin{equation}
 \label{eq::G}
G = \Big(\frac{e J^c_N}{\Delta\mu_N}\Big)_{\Delta T = 0} = \frac{L_{11}}{T},
\end{equation}
 \begin{equation}
 \label{eq::S}
 \mathcal{S} = -\Big(\frac{\Delta \mu_N}{e \, \Delta T}\Big)_{J^c_N = 0} = \frac{1}{T}\frac{L_{12}}{L_{11}}  ,
 \end{equation}
\begin{equation}
\label{eq::K}
K = \Big(\frac{J^h_N}{\Delta T} \Big)_{J^c_N = 0} = \frac{1}{T^2}\frac{L_{11} L_{22} -L_{21}L_{12}}{L_{11}}
\end{equation}
Importantly, we can express the efficiency for heat to work conversion with the standard two-terminal formula~\cite{Benenti2013} 
\begin{equation}
\eta=\frac{-X^\mu_N J^c_N}{J^h_N}=\frac{-L_{11} (X^\mu_N)^2 -L_{12}X^\mu_N X^T_P}{-L_{21}X^\mu_N -L_{22}X^T_P}. 
\end{equation}
One can also define the figure of merit $ZT = (G \mathcal{S}^2/K)\, T$ and the power factor $Q = G \mathcal{S}^2$. The 
former gives information about the maximum efficiency and the efficiency at maximum power~\cite{CurzonAhlborn1975, Broeck2005} $\eta(W_{\text{max}}) = 
(\eta_C/2)\, ZT/(ZT + 2)$, $\eta_C=1 - \Delta T/T$ being the Carnot efficiency, while the latter about the maximum power 
$W_{\text{max}} = Q (\Delta T)^2/4$. With these formulas the analogy between the SPN system and the two-terminal 
one is complete, allowing us to compare their performance.

\section{HSSC in hybrid  devices}
\label{sec:results}

As we shall show in this section, the heat-charge separation implemented through the SPN setup allows to control $G$ and $K$ separately. This will be at the origin of the enhancement of both the figure of merit $ZT$ and the power factor $Q$ with respect to the two-terminal setup.
We will use the Landauer-B\"uttiker scattering formalism~\cite{Buttiker1986,Landauer1957}, which is summarised in App.~\ref{app:LBsc} for multi-terminal hybrid superconducting systems.
We begin our analysis by considering low temperatures (within the Sommerfeld expansion) and studying a well-defined class of scattering matrices.
Quasiparticle transmission from the normal leads into the superconductor is exponentially suppressed and thus can be ignored. 
Thus the scattering probabilities entering Eq.~\eqref{eq::currents} just involve reservoirs N and P. 

In the following we will express the conductances (electrical and thermal) as well as the thermopower as functions of the parameters characterising the 
scattering matrix.
The aim is to sample this parameter space in order to make a statistical analysis of the thermoelectric performance. 
Assuming a single channel per spin per lead, in the Bogoliubov-de Gennes formalism (see App.~\ref{app:LBsc}) the total scattering matrix $\bf{S}_{tot}$ is $8 \times  8$. Supposing that there are no spin-mixing terms, it can be written in a diagonal block form
\begin{math}
\mathbf{S}_{tot}=
\begin{pmatrix}
\mathbf{S} & 0\\
0 & \mathbf{S'}
\end{pmatrix},
\end{math}
where the basis is $(c_{\uparrow,N},c_{\uparrow,P},c_{\downarrow,N}^\dagger,c_{\downarrow,P}^\dagger,c_{\uparrow,N}^\dagger,c_{\uparrow,P}^\dagger,c_{\downarrow,N},c_{\downarrow,P})$, where the operator $c_{\sigma, i}$ $(c^\dag_{\sigma,i})$ destroys (creates) an electron with spin $\sigma$ in lead $i = (N, S, P)$. The matrices $\mathbf{S}$ and $\mathbf{S'}$ are related by the particle-hole symmetry relations (see App.~\ref{app:LBsc}), so that assigning the elements of $\mathbf{S}$ is sufficient to know the whole $\mathbf{S}_{tot}$. 
For sake of simplicity we will consider symmetric unitary matrices.  A parametrization of such class of matrices is given by
\begin{equation}
\label{eq:SmatrixSPN}
{\bf S} =
\begin{pmatrix}
g_1 S_1 & g_2 S_2 \\
g_2 S_2^{\text{T}} & g_3 S_3 
\end{pmatrix},
\end{equation}
where $S_1$ and $S_3$ are $2\times 2$ symmetric unitary matrices, $S_2$ is a $2\times 2$ unitary matrix, $S_3=S_2^T S_1^* S_2$ and $g_1$, $g_2$, $g_3$ are such that the matrix
\begin{math}
\begin{pmatrix}
g_1 & g_2\\
g_2 & g_3
\end{pmatrix}
\end{math}
is unitary. For the sake of simplicity we assume the latter to be real, i.e. it can be written as 
$\begin{pmatrix}
g(E) & \sqrt{1-g(E)^2}  \\
\sqrt{1-g(E)^2}  & -g(E) 
\end{pmatrix},
$ 
where we made explicit the dependence on the energy $E$. Furthermore, we parametrize $S_1$ and $S_2$ as
\begin{equation}
S_1 = 
\begin{pmatrix}
- \rho_1(E) \e^{i(\theta_1 + 2 \beta_1)} & \sqrt{1-\rho_1(E)^2 }\e^{i \beta_1} \\
\sqrt{1-\rho_1(E)^2 }\e^{i \beta_1} & \rho_1(E) \e^{-i \theta_1 }
\end{pmatrix}
\end{equation}
and
\begin{equation}
S_2 = 
\begin{pmatrix}
- \rho_2(E) \e^{i(\theta_2 + \beta_2 + \gamma_2)} & \sqrt{1-\rho_2(E)^2 }\e^{i \beta_2} \\
\sqrt{1-\rho_2(E)^2 }\e^{i \gamma_2} & \rho_2(E) \e^{-i \theta_2}
\end{pmatrix},
\end{equation}
assuming the phases in the matrices to be energy independent. The last simplification that we impose is the following relation between the phases: $\beta_2 + \theta_2 = \beta_1 + \theta_1 + \frac{\pi}{2} $. Within this parametrization we assume $\theta_1$, $\beta_1$, $\theta_2$, $\beta_2$, and $\gamma_2$ to be real numbers and $g(E)$, $\rho_1(E)$ and $\rho_2(E)$ to be real functions of energy such that $0\le g(E), \, \rho_1(E), \, \rho_2(E) \le 1$ for any $E$.
With this notation and using the expressions for the Onsager coefficients in the Landauer-B\"uttiker approach given in App.~\ref{app:LBsc}, the Sommerfeld expansion yields the following transport coefficients [see Eqs.~\eqref{eq::G}-\eqref{eq::K}]:
\begin{widetext}
\begin{eqnarray}
\label{G_sommerfeld}
G &=&  8 - 8 g(0)^2 + \frac{8 (-1 + g(0)^2)^2}{-1 - \rho_2(0)^2 +  g(0)^2 (\rho_1(0)^2 + (3 - 2 \rho_1(0)^2) \rho_2(0)^2 + 2 (-1 + \rho_1(0)^2) \rho_2(0)^4)},   \\
\label{S_sommerfeld}
\mathcal{S} &=&2 \frac{\pi^2 T g(0) \rho_2(0)  \big[-\rho_2(0)(-1 + \rho_2(0)^2) \big[(-1 + \rho_1(0)^2) g'(0) + g(0) \rho_1(0) \rho_1'(0)\big] }{-3 \rho_2(0)^2 + 3 g(0)^2 (-1 + \rho_1(0)^2 + (3 - 2 \rho_1(0)^2) \rho_2(0)^2 + 2 (-1 + \rho_1(0)^2) \rho_2(0)^4)} \nonumber \\
&-& \frac{ g(0) (-1 + \rho_1(0)^2) (-1 + 2 \rho_2(0)^2) \rho_2'(0) \big]}{-3 \rho_2(0)^2 + 3 g(0)^2 (-1 + \rho_1(0)^2 + (3 - 2 \rho_1(0)^2) \rho_2(0)^2 + 2 (-1 + \rho_1(0)^2) \rho_2(0)^4)},\\
\label{K_sommerfeld}
K &= &- \frac{2 \pi^2 T}{3} \bigg[  -1 + \rho_2(0)^2 + g(0)^2 \big(\rho_2(0)^2 - 2 \rho_2(0)^4 + \rho_1(0)^2 \big(1 - 2 \rho_2(0)^2 + 2 \rho_2(0)^4 \big)\big) \bigg],
\end{eqnarray}
\end{widetext}
where the primed quantities are derivatives with respect to energy.
After the choices we made, we are left with six parameters [namely, $\rho_1(0)$, $\rho_1'(0)$, $\rho_2(0)$, $\rho_2'(0)$, $g(0)$, $g'(0)$] to 
control $G$, $\mathcal{S}$ and $K$.
We stress that we do not impose time reversal symmetry on the scattering matrix $\mathbf{S}$, Eq.~\eqref{eq:SmatrixSPN}, although our parametrization gives rise to a symmetric Onsager matrix.~\cite{notaF}

At this point we would like to draw the attention to the fact that the transport coefficients $G$, $S$ and $K$ are independent in a parameter region that is defined by the constraints imposed by the unitarity of the scattering matrix and from the Sommerfeld expansion. The independence of the transport coefficients can be appreciated from the way the six parameters, needed to parametrize the scattering matrix, enter Eqs.~\eqref{G_sommerfeld}- \eqref{K_sommerfeld}.
Indeed, if the value of $K$ in Eq.~\eqref{K_sommerfeld} is fixed, the value of $G$ given by Eq.~\eqref{G_sommerfeld} is not automatically determined, but instead it can be controlled by exploiting the other parameters. The same applies to $\mathcal{S}$ when $G$ and $K$ are fixed. 

We shall now discuss how the performance of the SPN system depends on these parameters. It is indeed important to 
verify if: i) HCCS is an advantage for thermoelectric conversion, ii) in the regime of HCCS the enhancement of the performance is generic or it requires additional fine tuning.
In order to assess the above issues we will first analyse to which extent heat and charge can be controlled independently and then we will 
study the figure of merit as a function of the parameters characterising the scattering matrix. Since we have to deal with six free parameters our 
analysis will be of statistical nature.

\begin{figure}[!h]
\centering
\includegraphics[width=\columnwidth,  keepaspectratio]{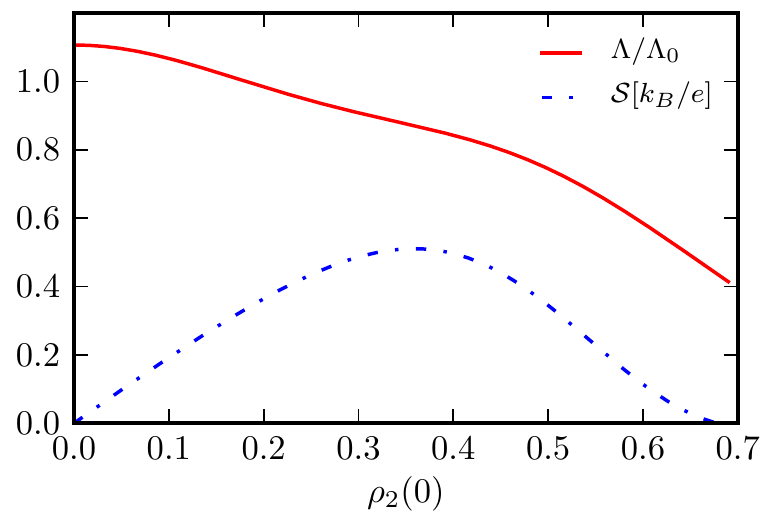}
\caption{Plot of the ratio $\Lambda/\Lambda_0$ and thermopower $\mathcal{S}$ for fixed $K = 10 \, (3 k_B T)/\pi^2 h$ as a function of $\rho_2(0)$. Here we show that using only one parameter [$\rho_2(0)$] we cannot control separately the thermopower and the ratio $\Lambda = K/(G T)$. The other parameters are: $\rho_1(0) = 0.8$, $g'(0) = 0.001(k_B T)^{-1}$, $\rho'_1(0) = 0.03(k_B T)^{-1}$ and $\rho'_2(0) = 0.3(k_B T)^{-1}$. } 
\label{fig:S}
\end{figure}
\subsection{Control of heat and charge currents}
\label{sec:violation}

Our strategy to test our ability to control the currents in the three-terminal device is to use, in the same spirit as in the Wiedemann-Franz law, the ratio between heat and electrical conductances. 
Using Eqs.~\eqref{G_sommerfeld} and~\eqref{K_sommerfeld} we can now calculate $ \Lambda = K/(G T)$ which can be seen by inspection not to be a constant, hence violating the Wiedemann-Franz law. Note that both $G$ and $K$ depend only on the coefficients $\rho_1(0), \, \rho_2(0)$, and $g(0)$.
In Fig.~\ref{fig:S} we plot, for a fixed value of $K$, the dimensionless ratio $\Lambda/\Lambda_0$ and the thermopower $\mathcal{S}$ as functions of the parameter $\rho_2(0)$.
More precisely, after fixing $K$ we extract  from Eq.~\eqref{K_sommerfeld} the parameter $g(0)$ which is a function of $K, \rho_1(0), \, \rho_2(0)$ and substitute it into Eq.~\eqref{G_sommerfeld}. Moreover, we impose the condition that the next order in the Sommerfeld expansion is much smaller than the one we take into account, restricting the range of admissible values of the other parameters (e.g. $\rho_2(0)$ can at most be $0.7$). 
For applying this condition we have to specify the values of the derivatives $g'(0)$, $\rho'_1(0)$ and $\rho'_2(0)$ even though they do not appear in Eqs.~\eqref{G_sommerfeld} and~\eqref{K_sommerfeld}. We assume higher order derivatives to be zero for simplicity. 
\begin{figure}[htb]
\centering
\includegraphics[width=\columnwidth,  keepaspectratio]{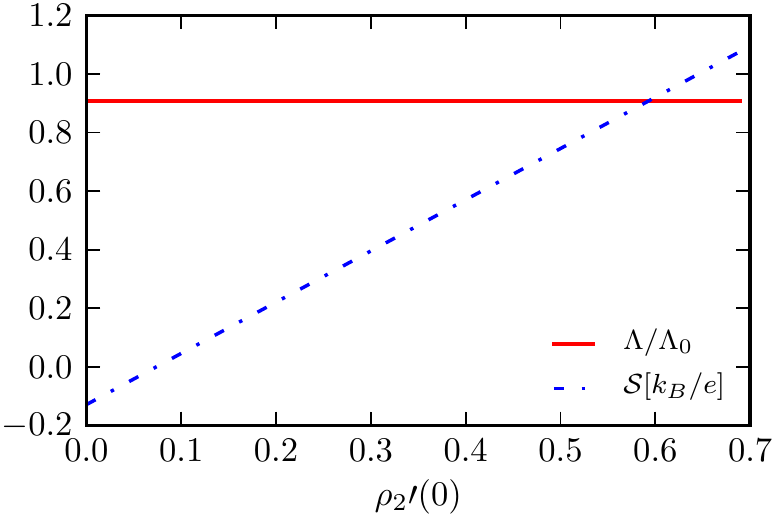}
\caption{Thermopower $\mathcal{S}$ from Eq.~\eqref{S_sommerfeld}, with $K = 10 \, (3 k_B T)/\pi^2 h$ as a function of $\rho'_2(0)$ [in units of $(k_B T)^{-1}$]. Using the additional degrees of freedom provided by the derivatives of the parameters [here we use $\rho_2'(0)$], we gain the control of the thermopower $\mathcal{S}$ without affecting the ratio $\Lambda$. The other parameters are: $\rho_1(0) = 0.8$, $g'(0) = 0.001(k_B T)^{-1}$, and $\rho'_1(0) = 0.1(k_B T)^{-1}$. }
\label{fig:S2} 
\end{figure}
The plot shows that $\Lambda$ is not a constant, but can be controlled by properly tuning the parameters of the scattering matrix.
Moreover, Fig.~\ref{fig:S} shows that $\mathcal{S}$ changes by varying $\rho_2(0)$ for fixed $K$.
The controllability of the transport coefficients can be further increased by fixing the values of both $K$ and $\Lambda$ using the parameters $\rho_1(0)$ and $\rho_2(0)$, and tuning the derivatives to change the thermopower. This is shown in Fig.~\ref{fig:S2} where $\Lambda$ and $\mathcal{S}$ are plotted as a function of $\rho'_2(0)$. Notably $\mathcal{S}$ spans a quite large interval of values, even changing sign. 
We conclude that the SPN system allows independent control of $G$, $K$ and $\mathcal{S}$.
This enhanced control is at the basis of the better performance that we are going to describe in the next Section.

\subsection{Thermoelectric performance}
\label{sec:performance}

In this section we compare on a statistical ground the thermoelectric performance of the SPN system with that of a generic two-terminal normal system by randomly generating the parameters of the scattering matrices and calculating the corresponding power factor $Q$ and figure of merit $ZT$.
\begin{figure}[htb]
\centering
\includegraphics[width=\columnwidth]{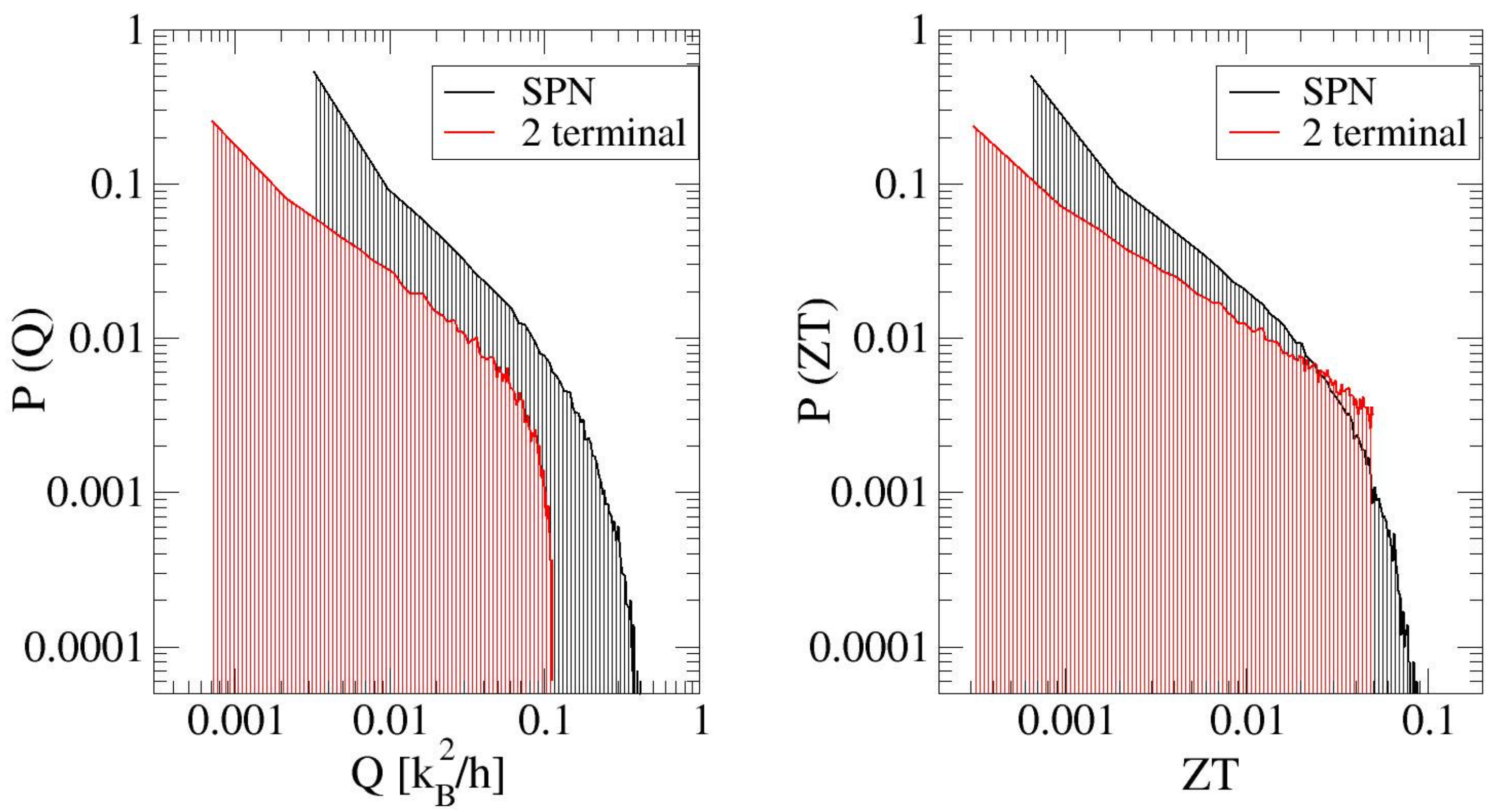}
\caption{(left) Probability histogram of the power factor $Q = G \mathcal{S}^2$ [in units of $k_B^2/h$] for the SPN system (black curve) and for the corresponding normal two-terminal system (red curve). The maximum value of $Q$ for the SPN setup is about 0.5 $k_B^2/h$, while it is about 0.2 $k_B^2/h$ for the two-terminal case. (right) Probability histogram of the figure of merit $ZT$ for the SPN system (black curve) and for the corresponding normal two-terminal case (red curve). The maximum of the SPN setup is just above 0.1, while it is about 0.05 for the two-terminal system.} 
\label{fig:histo1}
\end{figure}
Within the low temperature limit (Sommerfeld expansion), we perform a numerical simulation generating the parameters of the scattering matrix, of both the two-terminal and the SPN systems.
Such parameters are picked within a uniform distribution in the allowed ranges given by the conditions imposed by the unitarity of the scattering matrix~\cite{notaS} and the Sommerfeld expansion.
In Fig.~\ref{fig:histo1} we plot the probability of occurrence of a certain value of $Q$ (left panel) and $ZT$ (right panel). The plot shows that the SPN system (black histograms) has better performance than the normal two-terminal system (red histograms) for both the power factor $Q$ and the figure of merit $ZT$.
Indeed, the maximum value of $Q$ for the SPN system is about 0.5 $k_B^2/h$, while it is about 0.2 $k_B^2/h$ for the two-terminal one. The maximum of $ZT$ for the SPN system is just above 0.1, while it is about 0.05 for the two-terminal system. In Fig.~\ref{fig:histo2} we plot the correlations between $Q$ and $ZT$ for the same random data.
Each point in the plot corresponds to a particular realization of the scattering matrix of the two-terminal or the SPN system, for which the power factor and the figure of merit are calculated. 

\begin{figure}[!ht]
\centering
\includegraphics[width=\columnwidth]{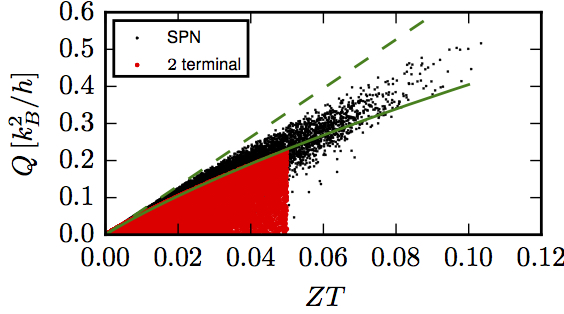}
\caption{Correlation between the values of $Q$ and $ZT$ for the same data as for Fig.~\ref{fig:histo1}: each point corresponds to a given random realisation. Red (black) points are relative to the two-terminal (SPN) setup. The green dashed curve represents the bound of Eq.~\eqref{eq::unitarybound}, that holds for both the SPN and the two-terminal system. The green solid curve, instead, represents the bound of Eq.~\eqref{eq::unitarybound2}, that is the stronger bound given by the unitarity on the two-terminal system.} 
\label{fig:histo2}
\end{figure}
Fig.~\ref{fig:histo2} shows that the distribution of points presents a triangular-like shape for the two-terminal setup.
An upper bound on the power factor $Q$ is given by the unitarity of the scattering matrix.
In fact, $Q$ and $ZT$ are related by the thermal conductance as $Q = ( K /T) ZT $ and, under the Sommerfeld expansion, $K$ is proportional to the probability of transmission of an electron from lead N to P, which cannot exceed unity.
This yields an upper bound given by 
\begin{equation}
\label{eq::unitarybound}
Q \leq \frac{2\pi^2k_B^2 ZT}{3h}.
\end{equation}
This is actually true for both the two-terminal and the SPN systems. For the two-terminal system a stronger bound, which produces the curvature of the 
upper side of the ``triangle", is given by the constraint that must be imposed on the derivative of the transmission amplitude with respect to energy imposed by unitarity of the scattering matrix~\cite{notaS}.
This implies the following expression for the maximum of $Q$
\begin{equation}
\label{eq::unitarybound2}
Q_{\text{max}}=\frac{2 e^2}{h} \Lambda_0 ZT \left(1+\frac{c}{e T} \sqrt{\frac{ZT}{\Lambda_0^2}}\right)^{-2},
\end{equation}
where $c$ is a given energy scale of the order of $k_B T$.
Furthermore, for the two-terminal system the power factor $Q$ can take all the values between $0$ and $Q_{\text{max}}$, thus filling the red ``triangle" of Fig.~\ref{fig:histo2}.
On the other hand, in the case of the SPN system the points are concentrated just below the line of the maximum. 
This is due to the fact that the value of $K/T$, given by Eq.~\eqref{K_sommerfeld}, cannot take all the values between $0$ and $2\pi^2k_B^2/(3h)$ because of the constraints imposed on, and the relations between, the parameters appearing in the expression.

The bound on the maximum value of $ZT$, instead, is given by the conditions on the higher orders of Sommerfeld expansion, that here we impose to be at least $10$ times smaller that the leading orders for both the two-terminal and the SPN system.
It is interesting to notice that for the SPN system the points with the highest power factor $Q$ are also the points with the highest figure of merit $ZT$: the maximum power automatically gives the maximum efficiency!
In particular the points with the best thermoelectric performance for the SPN system roughly correspond to the following values of the scattering probabilities: normal reflection in lead N, $R  \simeq 0$; normal transmission from lead N to lead P, $\mathcal{T} \simeq \frac{1}{4}$; Andreev reflection in lead N, $R_A   \simeq \frac{3}{16}$; and Andreev transmission from lead N to lead P, $\mathcal{T}_A \simeq \frac{9}{16}$.

\subsection{Coupled QDs in the SPN setup}
\label{sec:dots}

\begin{figure}[htb]
\centering
\includegraphics[width = \columnwidth]{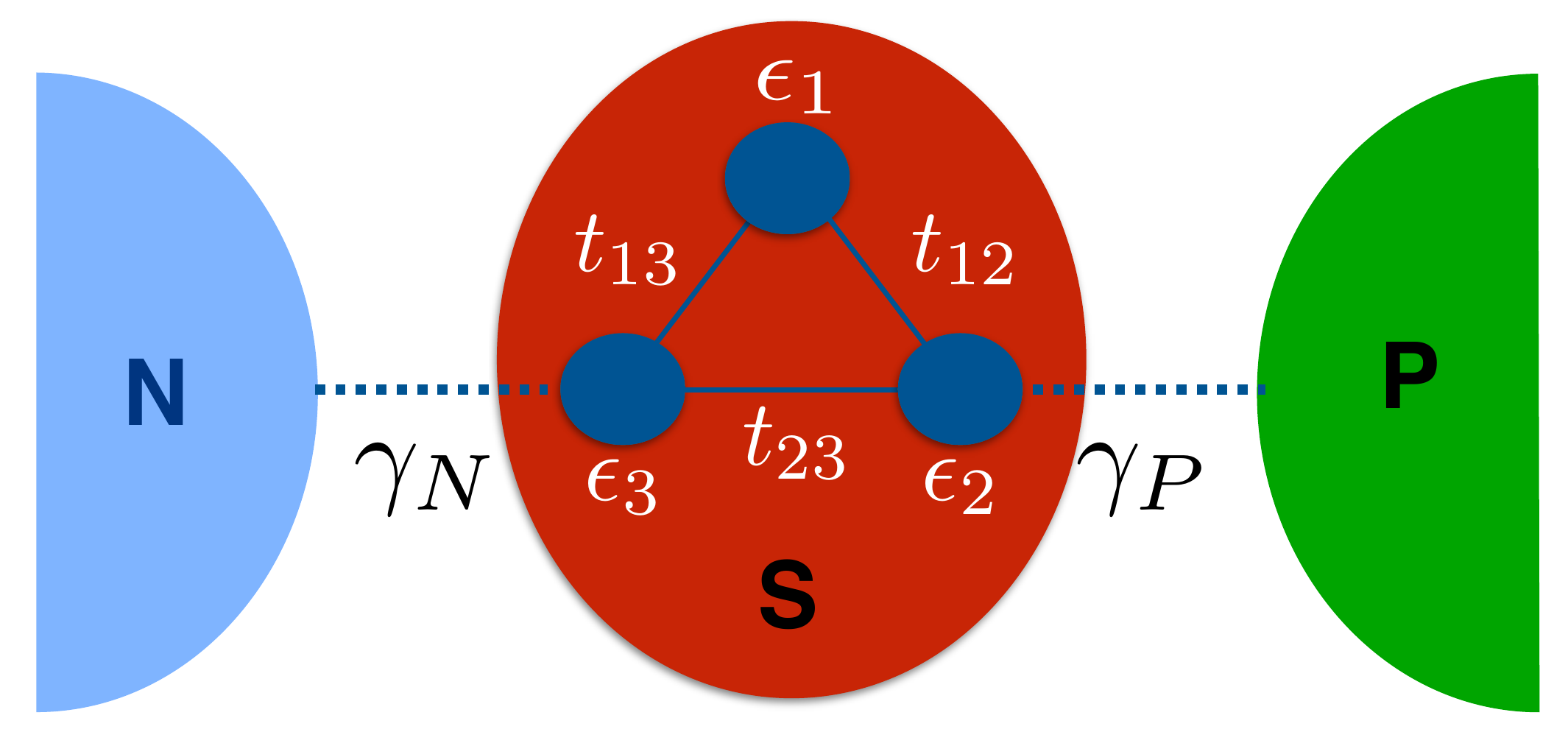}
\caption{A three coupled QDs (``tridot") system in the SPN setup. $\gamma_N$ and $\gamma_P$ label the coupling to the reservoirs N and P, respectively. The three QDs are coupled to a superconducting lead (S), with a fixed chemical potential.} 
\label{fig:SPNrandom3x3}
\end{figure}

We complete our analysis by assessing the thermoelectric performance of a specific SPN system composed of three coupled single-level (non-interacting) QDs (``tridot") connected to a normal lead (N), a voltage probe (P) and a superconducting lead (S), see Fig.~\ref{fig:SPNrandom3x3}.
The Hamiltonian describing the ``tridot" reads
\begin{equation}
\label{eq:hamiltonianRMT}
H = 
\begin{pmatrix}
H_e & \mathbb{1}\Delta  \\
\mathbb{1}\Delta^* & H_h
\end{pmatrix}
\end{equation}
where $H_e$ is the Hamiltonian relative to the electrons degree of freedom, and is given by
\begin{equation}
\label{He}
H_e = 
\begin{pmatrix}
\epsilon_1 & t_{12} & t_{13}\\
t_{12}^* & \epsilon_2 & t_{23} \\
t_{13}^* & t_{23}^* & \epsilon_3
\end{pmatrix},
\end{equation}
where $\epsilon_i$, $i = (1,2,3)$, is the onsite energy of the $i-$th dot, while $t_{ij}$, $\{i,j\} = (1,2,3)$, is the coupling between dot $i-$th and dot $j-$th.
Notice that $H_h = -H_e^*$ is the Hamiltonian relative to the holes, while $\Delta = 100$ $k_BT$ is the superconducting gap.
Note that the presence of the S lead is introduced in an effective way, whereby superconductivity is directly included in the Hamiltonian of the ``tridot".
The crucial point is that the superconductor chemical potential $\mu$ is fixed.
Furthermore, for simplicity, we have assumed that the superconducting pairing for all the QDs is equal as though originating from the fact that all QDs are equally coupled to the S lead.
In Fig.~\ref{fig:SPNrandom3x3}, $\gamma_N$ and $\gamma_P$ are the coupling energies to the N and P lead, respectively.

In order to make a statistical analysis we focus on random Hamiltonians, for the electron sector $H_e$, drawn from the Gaussian Orthogonal Ensemble (GOE) and from the Gaussian Unitary Ensemble (GUE). The former describe complex physical systems with time reversal symmetry (TRS), while the latter describe complex systems with broken TRS~\cite{Benakker1997, Haake2000,Benenti2011}. The TRS breaking is encoded in the complex part of the Hamiltonian~\eqref{He}. However this does not imply the spin degeneracy breaking, since we can apply a small Aharanov-Bohm flux through the plane of the ``tridot" (that breaks TRS), without a Zeeman component (that would have removed the spin degeneracy).
In both cases (with and without TRS), the elements of $H_e$ are drawn from a Gaussian probability distribution $\mathcal{N}(x, \Delta x)$, where $x$ is the mean and $\Delta x$ is the variance.
We use distributions with different mean for the diagonal elements (QD energy levels) and for the off diagonal elements (couplings between QDs), in order to have a band shift. The former are drawn from a Gaussian probability distribution $\mathcal{N}(k_BT, 10^3 k_BT)$ (mean equal to $k_BT$, and variance equal to $10^3 k_BT$). The latter are drawn from a Gaussian probability distribution $\mathcal{N}(0, 10^3 k_BT)$. The variance is chosen in order to obtain a smooth energy profile for the transmission probabilities and to be under the Sommerfeld expansion. 
\begin{figure}[!h]
\centering
\subfigure[][]{\label{fig:random3x3GOE}
\includegraphics[width=\columnwidth, keepaspectratio]{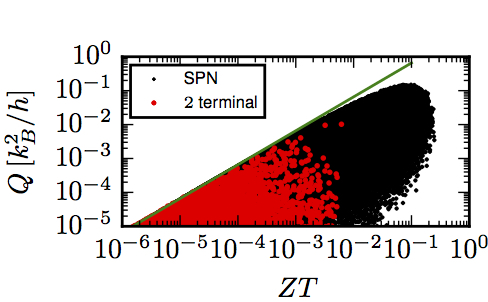}} \\
\subfigure[][]{\label{fig:random3x3GUE}
\includegraphics[width=\columnwidth, keepaspectratio]{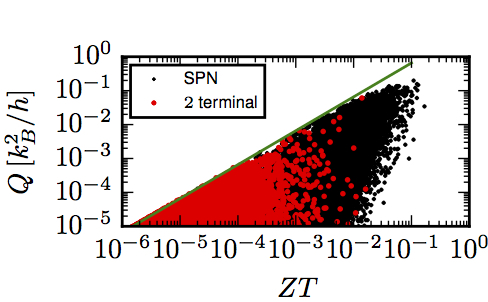}}
\caption{Correlation between the power factor $Q = G \mathcal{S}^2$ and the figure of merit $ZT$ relative to ``tridot" systems for the SPN setup (black points) and the two-terminal setup (red points). The green curve corresponds to the bound of Eq.~\eqref{eq::unitarybound}, given by the unitarity of the scattering matrix and set a maximum value for $Q$ as a function of $ZT$. In panel (a) we show the correlation for the Gaussian Orthogonal Ensemble (GOE), while in panel (b) for the Gaussian Unitary Ensemble (GUE). 
Both plots show that for the SPN setup both $Q$ and $ZT$ are one order of magnitude larger with respect to the corresponding values for the two-terminal system. Moreover it is possible to see that the increase of the performance is not due to the breaking of the TRS. The plot refers to $10^5$~Hamiltonian realizations, taking $\gamma_N=\gamma_P= \gamma = 10^3$~$k_BT$.} 
\label{fig:random3x3}
\end{figure}
From the Hamiltonian~(\ref{eq:hamiltonianRMT}) the Green function of the system is calculated for fixed values of $\gamma_N$ and $\gamma_P$.
The $4\times 4$ scattering matrix (there is spin degeneracy) is finally calculated from the Green function using the Fisher-Lee relation~\cite{Datta1995}.
Note that the two-terminal system, to be compared with the SPN setup, is simply described by the Hamiltonian $H_e$.
For each random realisation of the Hamiltonian of Eq.~\eqref{eq:hamiltonianRMT} we compute the power factor $Q$ and the figure of merit $ZT$.
The results are shown in Fig.~\ref{fig:random3x3}, in both the panels each point in the $Q-ZT$ plane represents a single realisation.
We notice that both $Q$ and $ZT$ are one order of magnitude larger in the SPN case with respect to the two-terminal configuration for both the GOE Hamiltonian in Fig.~\ref{fig:random3x3GOE}, and the GUE Hamiltonian in Fig.~\ref{fig:random3x3GUE}. 
This shows a significant enhancement, on a statistical ground, in the performance of the SPN with respect to the two-terminal system (the parameters of the system that realise the maxima of $Q$ and $ZT$ for both the two-terminal and the SPN system are given in App.~\ref{app:maxQ_ZT}).
The distribution of the points is similar to that of Fig.~\ref{fig:histo2}. Since for the ``tridot'' model the allowed values of $ZT$ (under Sommerfeld expansion) are much smaller than the values that we obtained from the model in the previous section (see Fig.~\ref{fig:histo2}), the bound on $Q$ of Eq.~\eqref{eq::unitarybound2} reduces to that of Eq.~\eqref{eq::unitarybound}. Instead, the bound on $ZT$ is again set by the Sommerfeld approximation. 
Here we notice that for the two-terminal system we cannot see the bound given in Eq.~\eqref{eq::unitarybound2} because of the small values of $ZT$.
As the temperature increases and the Sommerfeld expansion loses its validity, we have observed that the performance (for both $Q$ and $ZT$) of the SPN setup gets worse, eventually becoming comparable to that of the corresponding two-terminal setup.

\section{Conclusions}
\label{conclusions}

In this work we have analyzed the performance of a thermal machine which, by involving three reservoirs, allows for the implementation of a spatial separation between heat and charge currents in linear response.
Such machine can be naturally realised by connecting a conductor to a superconducting lead, a voltage probe and a normal lead (SPN system). 
Interestingly, the linear-response transport equations, written in terms of the Onsager matrix, turn out to be formally equal to those of a two-terminal conventional system.
Using this property we have made a comparison between the performance of these
two thermal machines in terms of the power factor $Q$ (that controls the maximum extracted power), and the figure of merit $ZT$ (that controls the efficiency at maximum power and the maximum efficiency).
Within the scattering approach we have described the SPN system with a parametrised scattering matrix.
We have shown that in the low temperature limit (where the Sommerfeld expansion holds) the SPN system violates the Wiedemann-Franz law and allows, to some extent, an independent control of electrical conductance, thermal conductance and thermopower ({\it i.e.} of heat and charge currents).
To assess the consequences of this on the thermoelectric performance of the SPN system we have made a statistical analysis by taking random values, over a uniform distribution, of the parameters contained in the scattering matrix.
We have thus shown, on statistical grounds, that the SPN system exhibits much larger values of $Q$ and $ZT$ with respect to the two-terminal counterpart. 
Further improvements (more than one order of magnitude) of the thermoelectric performance of the SPN setup has been confirmed on a specific physical system composed of three coupled quantum dots.
We believe that our results can be relevant in the experimental activity on thermoelectricity of nanoscale structures, which are typically conducted at low temperatures.

\section*{Acknowledgements}

This work has been supported by the EU project ``ThermiQ", by MIUR-PRIN ``Collective quantum phenomena: from strongly correlated systems to quantum simulators", by the EU project COST Action MP1209 ``Thermodynamics in the quantum regime", by the EU project COST Action MP1201 ``Nanoscale Superconductivity: Novel Functionalities through Optimized Confinement of Condensate and Fields", and by MIUR-FIRB2013 - Project Coca (Grant No.~RBFR1379UX).

\appendix

\section{Voltage- and temperature-probe setup}
\label{app-VP}

By imposing the thermal probe condition $J^h_N=0$ on reservoir N and the voltage probe condition $J^c_P=0$ on reservoir P, one can solve 
the second and third rows of Eq.~\eqref{eq::onsag_full} for the two biases $X^T_N$ and $X^\mu_P$, that will not depend directly on the currents 
$J^c_N$ and $J^h_P$, and find:
\begin{equation}\label{eq::manip_2probes}
\begin{pmatrix}
X^T_N \\
X^\mu_P 
\end{pmatrix}=-
\begin{pmatrix}
L_{22} &L_{23}\\
L_{32} & L_{33} 
\end{pmatrix}^{-1}
\begin{pmatrix}
L_{21}  &L_{24}\\
L_{31} & L_{34} 
\end{pmatrix}
\begin{pmatrix}
X^\mu_N \\
X^T_P 
\end{pmatrix}.
\end{equation}
From the first and forth rows of Eq.~\eqref{eq::onsag_full} one can define an effective two-terminal-like Onsager matrix, with
\begin{equation}\label{eq::onsag_2probes}
\begin{pmatrix}
J^c_N \\
J^h_P
\end{pmatrix}
= 
\begin{pmatrix}
L'_{11} & L'_{12} \\
L'_{21} & L'_{22} 
\end{pmatrix}
\begin{pmatrix}
X^\mu_N \\
X^T_P
\end{pmatrix},
\end{equation}
where the primed Onsager coefficients are obtained by substituting the expressions of Eq.~\eqref{eq::manip_2probes} into 
Eq.~\eqref{eq::onsag_full}.  What has been given above should be considered as a definition of HCCS and not a way to implement it.

\section{Landauer-B\"uttiker formalism for hybrid superconducting systems}
\label{app:LBsc}

Let us consider a mesoscopic system composed of a conductor to which $n>1$ leads are attached. Each lead is in equilibrium with a fermionic reservoir to which a Fermi distribution function is associated, so that a lead is characterized by a temperature and a chemical potential.  At energy $E$ the i-th lead has $N_i(E)$ open transverse channels. We allow the possibility to have superconductivity in the system and for simplicity we describe it using the Bogoliubov-de Gennes (BdG) formalism~\cite{BdG}, which doubles the degrees of freedom by introducing ``hole" states. The BdG Hamiltonian is particle-hole symmetric by construction, i.e. it is such that $\{H_{BdG},\mathcal{C}\}=0$, where $\mathcal{C}$ is the charge-conjugation operator and the curly parentheses stand for the anti-commutator. A hole state is the charge-conjugate of an electronic state, e.g. if the operator $c_{k,\sigma}$ destroys an electron of momentum $k$ and spin $\sigma$, the operator $\mathcal{C}c_{k,\sigma}\mathcal{C}=c^\dagger_{k,\sigma}$ destroys a hole of momentum $k$ and spin $\sigma$. For completeness we mention that $\mathcal{C}$ is a anti-unitary operator hence besides exchanging creation and annihilation operators one must take the complex conjugate of the numeric coefficients.

Particle-hole symmetry implies that the occupation of a hole state is the complementary to the occupation of an electronic state with opposite energy:
\begin{equation}\label{eq::phs_fermi}
f^-_j(E)=1-f^+_j(-E) ,
\end{equation} 
where $f^-_j$ is the distribution function for a hole in lead $j$ and $f^+_j$ is the analogous for electrons. We can then write a generalized expression for the Fermi distribution function as follows:
\begin{equation}\label{eq::fermi_distribution}
f_j^\alpha(E) = \frac{1}{1+\exp\big[\beta_j (E-\alpha(\mu_j-\mu_s) ) \big] },
\end{equation}
where $\mu_j$ is the chemical potential of the $j-$th lead, $\mu_s$ is the chemical potential of the superconductors which we take as a reference for the energies, $\beta_j = (k_B T_j)^{-1}$ is the inverse temperature of the $j$-th lead and $\alpha$ is equal to $+$ for electrons and $-$ for holes.
Assuming coherent transport in the conductor, one can express the charge and energy currents flowing through the normal leads in terms of scattering probabilities using the Landauer-B\"uttiker formalism generalized to include superconductivity:
\begin{widetext}
\begin{equation}\label{eq::currents}
\begin{aligned}
 J^c_i &= \sum_j^{n}\Big(-\frac{e}{h}\Big) \sum_{\alpha\sigma\beta\sigma'} \alpha \int_0^{+\infty} dE \, P_{ij}^{\alpha\sigma\beta\sigma'}(E) f^\beta_j(E) + \frac{e}{h}\sum_{\alpha\sigma} \alpha \int_0^{+\infty} dE \, N_i^{\alpha\sigma}(E)f^\alpha_i(E),\\
 J^{u}_i &= \sum_j^{n}\Big(-\frac{1}{h}\Big) \sum_{\alpha\sigma\beta\sigma'}  \int_0^{+\infty} dE \, (E + \alpha \mu_s) P_{ij}^{\alpha\sigma\beta\sigma'}(E) f^\beta_j(E) +  \frac{1}{h}\sum_{\alpha\sigma} \int_0^{+\infty} dE \,(E + \alpha \mu_s) N_i^{\alpha\sigma}(E)f^\alpha_i(E),
\end{aligned}
\end{equation}
\end{widetext}
where $J_i^c$ is the charge current in the i-th lead, $J^u_i$ is the energy current in the i-th lead, $e$ is the electron charge, $h$ is the Planck constant and $N_i^{\alpha \sigma}(E)$ is the number of open channels at energy $E$ for particles of type $\alpha$ and spin $\sigma$. In Eq.~\eqref{eq::currents} $P_{ij}^{\alpha\sigma\beta\sigma'}(E)$ is the probability for a particle of type $\beta$, spin $\sigma'$ and energy $E$ incoming from lead $j$ to be elastically scattered as a particle of type $\alpha$ and spin $\sigma$ into the $i$-th lead. 
The probability of scattering is related to the scattering matrix by $P_{ij}^{\alpha\sigma\beta\sigma'}(E)=\sum_{a,b}|S^{\alpha\sigma,
\beta\sigma'}_{(i,a),(j,b)}(E)|^2$,
where $a$ and $b$ are the transverse channels in lead $i$ and $j$ respectively.
To avoid double-counting that would have been introduced by the BdG formalism, the integrals over the energies run from $0$ to $+\infty$ instead of starting from $-\infty$. Here zero energy corresponds to the Fermi energy of the superconductors.
Due to particle-hole symmetry the probability of scattering from lead $i$ to lead $j$ satisfies the relation
\begin{equation}\label{eq::phs_probabilities}
P_{ij}^{\alpha\sigma,\beta\sigma'}(E)=P_{ij}^{-\alpha\sigma,-\beta\sigma'}(-E).
\end{equation}
The unitarity of the scattering matrix yields the following sum rules:
\begin{equation}\label{eq::unitarity_probability}
\sum_{j,\sigma',\beta}P_{ij}^{\alpha\sigma,\beta\sigma'}(E)=N_i^{\alpha,\sigma}, \qquad \sum_{i,\sigma,\alpha}P_{ij}^{\alpha\sigma,\beta\sigma'}(E)=N_j^{\beta,\sigma'}.
\end{equation}
The expressions of Eq.~\eqref{eq::currents} can be simplified by substituting Eqs.~\eqref{eq::phs_fermi},\eqref{eq::phs_probabilities} and~\eqref{eq::unitarity_probability} resulting in
\begin{widetext}
\begin{equation}\label{eq::currents_afterphs}
\begin{aligned}
& J^c_i = \frac{e}{h} \sum_{j\sigma\beta\sigma'} \int_{-\infty}^{+\infty} dE \, \Big[N_i^{+\sigma}(E) \delta_{ij} \delta_{\sigma \sigma'} \delta_{\beta+} - P_{ij}^{+\sigma\beta\sigma'}(E)\Big] f^\beta_j(E) ,\\
& J^{u}_i = \frac{1}{h} \sum_{j\sigma\beta\sigma'}  \int_{-\infty}^{+\infty} dE \, (E + \mu_s) \Big[N_i^{+\sigma}(E) \delta_{ij} \delta_{\sigma \sigma'} \delta_{\beta+} - P_{ij}^{+\sigma\beta\sigma'}(E)\Big] f^\beta_j(E).
\end{aligned}
\end{equation}
\end{widetext}
Once the charge and energy currents are determined in the normal leads, the sum of the currents in the superconducting leads $J^c_{SC}=\sum_j^{SC}J^c_j$ and $J^u_{SC}=\sum_j^{SC}J^u_j$ can be calculated exploiting Kirchhoff's sum rules $\sum_i J^c_i=0$ and $\sum_i J^u_i=0$, which are a consequence of charge and energy conservation.
From the first law of thermodynamics one can also define a heat current $J^Q_i$ in the i-th lead as
\begin{equation}\label{eq::heat_current}
J^h_i= J^u_i - \frac{\mu_i}{e} J^c_i .
\end{equation}
Let us remark that in a general case there is no sum rule for the heat currents. However, one can notice that in the superconducting leads the heat current is $J^h_i= J^u_i - \frac{\mu_s}{e} J^c_i$, hence the sum of the heat currents over the superconducting leads is $\sum_j^{SC}J^h_j= J^h_{SC}=J^u_{SC} - \frac{\mu_s}{e} J^c_{SC}$ which can be determined by Kirchhoff's sum rules on charge and energy currents.
A hybrid mesoscopic device can be thought as a thermal machine, for example for heat to work conversion.
One can define the efficiency of the thermal machine as the ratio between the work $W$ extracted from the engine when it absorbs heat $\mathcal{Q}$. By convention we assume positive the heat flowing into the system, hence in the steady state the definition of the efficiency is equivalent to
\begin{equation}\label{eq::eta}
\eta=\frac{\dot{W}}{\dot{\mathcal{Q}}}=\frac{\sum_i J^h_i}{\sum^+_i J^h_i}=\frac{-\sum_i \Delta \mu_i J^c_i}{e\sum^+_i J^h_i},
\end{equation}
where the dot indicates a derivative with respect to time and the apex + in the denominator means that the sum is restricted to positive heat currents. Time derivative of the work $\dot W$ must be positive for the machine to work as a heat to work converter, otherwise we are dealing with a refrigerator and the definition of $\eta$ is no longer valid.\\

Assuming small temperature- and voltage-biases, we can expand the Fermi distribution function of Eq.~\eqref{eq::fermi_distribution} at first order in such quantities:
\begin{align}\label{eq::fermi_taylor}
f^\alpha_j(E) \simeq f(E) + \frac{\partial f^\alpha_j}{\partial X^T_j}\biggr\rvert_{(E,T)}  X^T_j +  \frac{\partial f^\alpha_j}{\partial  X^\mu_j}\biggr\rvert_{(E,T)} X^\mu_j , \\
\frac{\partial f^\alpha_j}{\partial X^T_j}\biggr\rvert_{(E,T)}= -\frac{E}{T}\frac{\partial f}{\partial E}, \qquad \frac{\partial f^\alpha_j}{\partial  X^\mu_j}\biggr\rvert_{(E,T)}=- \alpha e \frac{\partial f}{\partial E} ,\nonumber
\end{align}
where we defined $f(E)=(1+\e^{\frac{E}{k_B T}})^{-1}$. Since the scattering matrix is independent of the biases, we can use Eq.~\eqref{eq::fermi_taylor} to linearise the currents of Eqs.~\eqref{eq::currents_afterphs} and \eqref{eq::heat_current} as follows:
\begin{equation}\label{eq::linearized_currents}
\begin{aligned}
J^c_i= \sum_j \mathcal{G}_{ij} \Delta V_j + \sum_j \mathcal{D}_{ij} \Delta T_j, \\
J^h_i= \sum_j \mathcal{M}_{ij} \Delta V_j + \sum_j \mathcal{K}_{ij} \Delta T_j,
\end{aligned}
\end{equation}
where we defined the quantities
\begin{widetext}
\begin{equation}\label{eq::onsager_coeff}
\begin{aligned}
\mathcal{G}_{ij}=\frac{e^2}{h} \sum_{\sigma\sigma'} \int_{-\infty}^{+\infty} dE \, \Big[N_i^{+\sigma}(E) \delta_{ij} \delta_{\sigma \sigma'} - P_{ij}^{+\sigma+\sigma'}(E) + P_{ij}^{+\sigma-\sigma'}(E)\Big] \Big(-\frac{\partial f}{\partial E}\Big), \\
\mathcal{D}_{ij}=\frac{e}{h} \sum_{\sigma\sigma'} \int_{-\infty}^{+\infty} dE \, \frac{E}{T} \Big[N_i^{+\sigma}(E) \delta_{ij} \delta_{\sigma \sigma'} - P_{ij}^{+\sigma+\sigma'}(E) - P_{ij}^{+\sigma-\sigma'}(E)\Big] \Big(-\frac{\partial f}{\partial E}\Big), \\
\mathcal{M}_{ij}=\frac{e}{h} \sum_{\sigma\sigma'} \int_{-\infty}^{+\infty} dE \, E \Big[N_i^{+\sigma}(E) \delta_{ij} \delta_{\sigma \sigma'} - P_{ij}^{+\sigma+\sigma'}(E) + P_{ij}^{+\sigma-\sigma'}(E)\Big] \Big(-\frac{\partial f}{\partial E}\Big), \\
\mathcal{K}_{ij}=\frac{1}{h} \sum_{\sigma\sigma'} \int_{-\infty}^{+\infty} dE \, \frac{E^2}{T} \Big[N_i^{+\sigma}(E) \delta_{ij} \delta_{\sigma \sigma'} - P_{ij}^{+\sigma+\sigma'}(E) - P_{ij}^{+\sigma-\sigma'}(E)\Big] \Big(-\frac{\partial f}{\partial E}\Big).
\end{aligned}
\end{equation}
\end{widetext}
Notice that the contributions from energies above the gap to the integrals defining the Onsager coefficients of Eqs.~\eqref{eq::onsager_coeff} are exponentially suppressed by the Fermi factor and will be neglected in the following. 
Let us notice that from the unitarity of the scattering matrix it follows that the diagonal coefficients are always positive or zero. 
The coefficients $\mathcal{G}_{ij}$, $\mathcal{D}_{ij}$, $\mathcal{M}_{ij}$ and $\mathcal{K}_{ij}$ are related to the usual Onsager coefficients\cite{Benenti2013} $L_{ij}$ via the following identifications:
\begin{equation}
\begin{pmatrix}
L_{11} & L_{12} & L_{13} & L_{14} \\
L_{21} & L_{22} & L_{23} & L_{24} \\
L_{31} & L_{32} & L_{33} & L_{34} \\
L_{41} & L_{42} & L_{43} & L_{44} \\
\end{pmatrix}
=T
\begin{pmatrix}
\mathcal{G}_{11} & T \mathcal{D}_{11} & \mathcal{G}_{12} &T \mathcal{D}_{12} \\
\mathcal{M}_{11} &T \mathcal{K}_{11} &  \mathcal{M}_{12} &T \mathcal{K}_{12} \\
\mathcal{G}_{21} &T \mathcal{D}_{21} & \mathcal{G}_{22} &T \mathcal{D}_{22} \\
\mathcal{M}_{21} &T \mathcal{K}_{21} & \mathcal{M}_{22} &T \mathcal{K}_{22} \\
\end{pmatrix}
\end{equation}

\section{Hamiltonians of the maxima of $Q$ and $ZT$ for the ``tridot system".}
\label{app:maxQ_ZT}

In this section we provide the parameters of the electronic Hamiltonians $H_e$ that realise the maximum values of $Q$ and $ZT$ for the two-terminal and SPN systems. 
The Hamiltonians that give the maxima for the Gaussian Orthogonal Ensemble are
\begin{equation}
H_e^{\text{2 term, GOE}} = 
\begin{pmatrix}
1.556 &  -0.983 & -0.046 \\
 -0.983 &  0.109  & 0.18 \\
 -0.0458 & 0.18 &  -0.001 
 \end{pmatrix} \gamma,
 \end{equation}
 
 \begin{equation}
 H_e^{\text{SPN, GOE}} = 
\begin{pmatrix}
-1.888 &  -1.074 &  0.209 \\
 -1.074 &  0.724 &  -0.834 \\
 0.209 &  -0.834 & 0.387
 \end{pmatrix}\gamma.
 \end{equation}

The Hamiltonians for the maxima for the Gaussian Unitary Ensemble are
\begin{widetext}
\begin{equation}
H_e^{\text{2 term, GUE}} = 
\begin{pmatrix}
0.729 & 0.583 + 2.375 \, i & 0.273 +0.115 \, i \\
0.583 - 2.375 \, i &  0.425 & 0.062 -0.226 \, i \\
0.273 -0.115 \, i &  0.062 + 0.226\, i & 0.039 
\end{pmatrix} \gamma,
\end{equation}
\end{widetext}
\begin{widetext}
\begin{equation}
H_e^{\text{SPN, GUE}} = 
\begin{pmatrix}
0.064 &  -0.923 + 0.484 \, i & -0.16 - 0.921\, i \\  
-0.923-0.484 \, i & 1.213 & -0.39 + 0.752 \, i \\        
-0.16+921.13\, i & -0.39 - 0.752 \, i & 0.481      
\end{pmatrix} \gamma,
\end{equation}
\end{widetext}
where all the energies are expressed in terms of the coupling energy $\gamma$. Note that, although the values of the Hamiltonian parameters are much greater than the temperature, the large value of the coupling energy $\gamma$ makes the linear response coefficients significantly affected by such parameters.
Indeed, the transmission profiles corresponding to these Hamiltonians turn out to be stretched on a (large) scale set by $\gamma$, and hence are almost flat within the transport window (that in the linear response regime is given by the width of the derivative of the Fermi function and is of the order of few $k_BT$).



\begin{thebibliography}{88}
\bibitem{notaF}
Note that spin-dependent effects are included in this parametrization.
\bibitem{notaS}
Unitarity requires that the functions $g(E)$, $\rho_1(E)$ and $\rho_2(E)$ are smaller than unity for any energy $E$.
This implies the following condition on their derivatives with respect to energy: $\tau'(E)\leq [1-\tau(E)]/c$, where $\tau=g,\rho_1,\rho_2$ and $c$ is an energy scale.
The same holds for the transmission amplitude of the two-terminal setup $t(E)$.
\end{thebibliography}
\end{document}